\documentclass{article}
\usepackage{spconf,amsmath,graphicx,hyperref}
\usepackage{xcolor}
\usepackage{booktabs,multirow}
\usepackage{amssymb}
\usepackage{arydshln}

\usepackage{enumitem}
\usepackage{bm}

\title{Lightweight and Perceptually-Guided Voice Conversion for Electro-Laryngeal Speech}
\name{Benedikt Mayrhofer$^{1,3}$, Franz Pernkopf$^1$, Philipp Aichinger$^{2,3}$, Martin Hagmüller$^{1,3}$}

\address{$^1$Signal Processing and Speech Communication Laboratory, Graz University of Technology\\ 
$^2$Department of Otorhinolaryngology, Div. Phoniatrics-Logopedics, Medical University of Vienna\\
$^3$Comprehensive Centre for AI in Medicine, Medical University of Vienna\\
{\{benedikt.mayrhofer, hagmueller, pernkopf\}@tugraz.at, philipp.aichinger@meduniwien.ac.at}}
\begin{document}
\ninept
\maketitle
\begin{abstract}
Electro-laryngeal (EL) speech is characterized by constant pitch, limited prosody, and mechanical noise, reducing naturalness and intelligibility. We propose a lightweight adaptation of the state-of-the-art StreamVC framework to this setting by removing pitch and energy modules and combining self-supervised pretraining with supervised fine-tuning on parallel EL \& healthy (HE) speech data, guided by perceptual and intelligibility losses. Objective and subjective evaluations across different loss configurations confirm their influence: the best model variant, based on WavLM features and human-feedback predictions (\textit{+WavLM+HF}) drastically reduces character error rate (CER) of EL inputs, raises naturalness mean opinion score (nMOS) from 1.1 to 3.3, and consistently narrows the gap to HE ground-truth speech in all evaluated metrics. These findings demonstrate the feasibility of adapting lightweight voice conversion (VC) architectures to EL voice rehabilitation while also identifying prosody generation and intelligibility improvements as the main remaining bottlenecks. 
\end{abstract}
\section{Introduction}
\label{sec:intro}

Nearly 200.000 new cases of laryngeal cancer are diagnosed worldwide each year~\cite{hoffmann2021laryngectomy}. Many affected patients need to undergo a total laryngectomy, losing their natural voice source. For voice rehabilitation, a significant number of laryngectomees rely on an electro-larynx device~\cite{kaye2017electrolarynx}. Most EL speech is characterized by constant, monotonic pitch, reduced prosody, and mechanical noise from the buzzer. These characteristics degrade speech naturalness, expressiveness and intelligibility~\cite{fuchs2016bionic}.
To improve EL speech, prior work has investigated several approaches, notably VC techniques~\cite{ma2023twostage, yang2023electrolaryngeal, ma2025elvc, mayrhofer2025maveba_vc}. VC aims to transform EL speech into a more natural voice by converting its acoustic features to those of HE speech.


Real-time VC of EL-to-HE speech has recently gained attention. Kobayashi \textit{et al.} \cite{kobayashi2023lowlatency, kobayashi2021lowlatency} proposed low-latency VC using causal CLDNN, enabling real-time processing. However they reported that this approach limits the modeling accuracy of acoustic features, reducing naturalness compared to HE speech.

Parallel advances in streaming VC for HE speech have shown that high-quality VC is possible with low-latency ~\cite{LLVC2023, wang2024streamvoiceplus}. In particular, StreamVC~\cite{yang2024streamvc} implements causal, any-to-any VC by combining self-supervised content units with SoundStream~\cite{zeghidour2022soundstream} encoder-decoder architectures and whitened fundamental frequency ($F_0$)/energy conditioning, achieving 70.8 ms latency on a Pixel 7 smartphone. However, its $F_0$ module is unsuitable for EL speech due to the absence of $F_0$ variation. It would thus produce flat output unless the model is modified to generate a pitch contour. Additionally EL speech has often a slower speaking rate and different acoustic characteristics than HE speech~\cite{fuchs2016bionic}, complicating alignment between source and target speech features.

In this paper, we introduce: (1)  a custom time alignment technique using Whisper~\cite{radford2022robust} Encoder Output features and dynamic time warping (DTW) for EL-HE speech pairs, (2) the adaptation of the state-of-the-art StreamVC model architecture and training paradigm to handle EL-to-HE VC, (3) a systematic comparison of different perceptual- and intelligibility-guided loss configurations to assess their impact on natural speech restoration within a lightweight model architecture, and (4) objective and subjective evaluations of our approach. Together, perceptual-guided losses proved to be critical for performance, the best loss configuration, based on WavLM~\cite{chen2022wavlm} features and human-feedback predictions (\textit{+WavLM+HF}) reduced CER and raised naturalness while outperforming EL speech and strong baselines (FreeVC, XVC). Audio samples\footnote{https://spsc-tugraz.github.io/lw-elvc-icassp26/} are publicly available.

\vspace{-1mm}
\section{Resources}

\subsection{Database}

For pretraining, three publicly available German speech corpora are used: Common Voice (v22.0)~\cite{ardila2019commonvoice}, HUI-Audio-Corpus-German~\cite{puchtler2021huiaudio}, and the German portion of the Multilingual LibriSpeech (MLS) corpus~\cite{pratap2020mls}. The Common Voice corpus provides crowd-sourced speech data with diverse recording conditions. To ensure quality, Common Voice utterances were DNSMOS Pro~\cite{cumlin2024dnsmos}-scored, and the top 10\% (192h) retained. To avoid speaker imbalance, HUI and MLS were limited to 10h per speaker, yielding 180h and 170h, respectively. In total, 542h of HE German speech were collected.

For fine-tuning and evaluation, we use the Austrian-German parallel ELHE database~\cite{Fuchs2024ELHE}, comprising up to 500 utterances recorded both as HE and with an EL device. The EL data were produced by HE speakers using an EL device, and supplemented with recordings from post-laryngectomy patients on YouTube. The final dataset includes 8 pseudo EL speakers, 8 real EL speakers, and 8 HE speakers. The corpus contains 2.75h each of HE and EL speech (5.5h total), partitioned 80–10–10\% into training, development, and validation.
\subsection{Model Architecture}
Our system adapts the architecture of StreamVC~\cite{yang2024streamvc}. The model is implemented as a fully causal convolutional network. This design makes the system suitable for real-time streaming VC.

\begin{figure}[t]
\vspace{-0.5mm}
  \centering
    \includegraphics[width=1.0\linewidth]{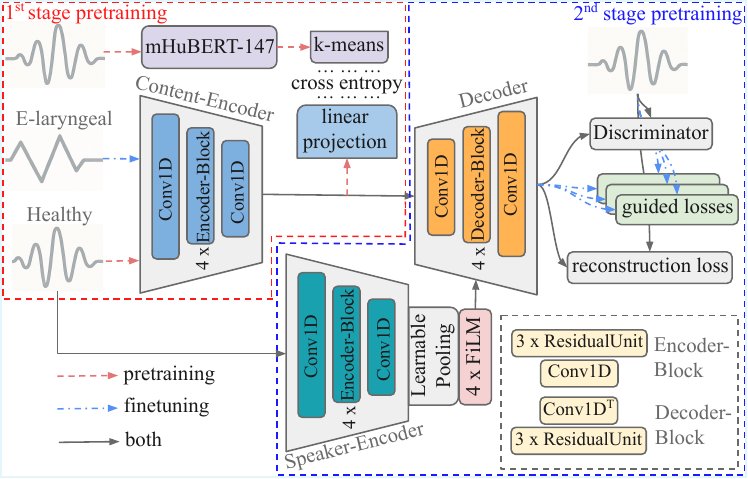}
    \vspace{-5.5mm}
  \caption{Adapted StreamVC architecture for EL-HE voice conversion. It consists of a content encoder, speaker encoder with learnable pooling, and FiLM-conditioned decoder. Training uses mel-spectral reconstruction, adversarial/feature-matching, and additional guided losses (perceptual, human-feedback, intelligibility, and $F_{0}$).}
  \label{fig:model}
  \vspace{-5mm}
\end{figure}

Fig.~\ref{fig:model} shows the architecture, it consists of three main modules: content encoder, speaker encoder with learnable pooling, and a decoder conditioned on the speaker embeddings via FiLM~\cite{perez2018film} conditioning layers. In contrast to StreamVC, we omit the pitch and energy estimation modules. This choice is motivated by two factors: (i) EL speech does not contain prosodic cues such as $F_0$ variations, and (ii) excluding these components reduces model complexity. We also replace StreamVC’s HuBERT-base teacher with a multilingual mHuBERT-147~\cite{zanonboito2024mhubert}  teacher for distillation, better suited to German speech. For Training we added different perceptually guided losses. The resulting system focuses on disentangling content from speaker information and reconstructing the HE waveform from EL input.

The model code is adopted from an unofficial implementation of StreamVC\footnote{https://github.com/yuval-reshef/StreamVC}; all other components were implemented by us. Waveform synthesis is trained in a GAN framework with a discriminator architecture following the HiFi-GAN~\cite{kong2020hifigan} MPD and MSD design. The final model has approximately 30M parameters (18M content-encoder, 4M speaker-encoder, and 8M decoder) and a total size of 123MB.

\vspace{-1mm}
\section{Training}

\subsection{Pretraining}
We first pretrain the model on HE speech in a self-supervised manner, where the input and target are identical waveforms. The objective is to disentangle speaker characteristics from linguistic content and reconstruct the waveform. During inference, speaker embeddings can be swapped to achieve VC. Since ELHE data is limited, direct training would not yield robust waveform synthesis. Pretraining on HE speech is done to establish general VC capability before fine-tuning on ELHE data.

Pretraining follows the two-stage setup of StreamVC. In the first stage, the content encoder is trained as a student network to predict discrete units derived from the teacher model. The HuBERT features are clustered into 100 units using k-means, and the student encoder is trained to classify these units using cross-entropy loss. In the second stage, the encoder is frozen to prevent speaker leakage, and the decoder is trained jointly with the speaker encoder to reconstruct the waveform. We first train for 150k steps with only mel-spectral reconstruction loss, and continue for 400k steps with additional adversarial and feature matching~\cite{yang2024streamvc} losses to achieve fast and stable convergence.

We use the Adam optimizer~\cite{kingma2015adam} with cosine decay scheduling. During pretraining, the model is trained with learning rates of $1.0e^{-4}$ (gen) and $2.0e^{-4}$ (disc), minimum rates of $2.5e10^{-5}$ (gen) and $1.25e10^{-5}$ (disc) and $\beta_{1}{=}0.5$, $\beta_{2}{=}0.9$, and a batch size of 16. The generator operates on 48k-sample segments (3\,s at 16 kHz), while the discriminator used 6400-sample segments (400\,ms). 

\subsection{Preprocessing of the parallel corpus}

\begin{figure}[t]
\vspace{-0.5mm}
  \centering
    \includegraphics[width=1.01\linewidth]{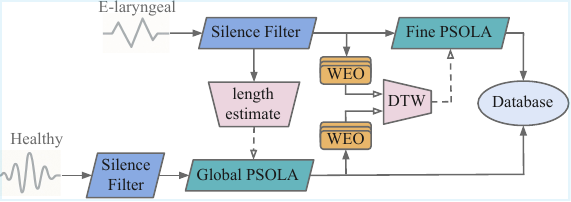}
    \vspace{-5.5mm}
  \caption{Time-alignment pipeline for EL--HE pairs: silence filter, PSOLA time-stretching, WEO feature alignment via DTW.}
  \label{fig:prepipeline}
  \vspace{-5mm}
\end{figure}

To enable supervised fine-tuning, the ELHE corpus is temporally aligned. Given the large acoustic mismatch between EL and HE speech, standard alignment approaches based on low-level acoustic features (e.g., mel-spectra or MFCC features) are found to be suboptimal. Instead, we fine-tune a Whisper-small model on the ELHE database. The Whisper Encoder Output (WEO) features serve as content-oriented feature base for alignment, providing shared representation across both speech domains. The preprocessing pipeline is shown in Fig. \ref{fig:prepipeline}. First, silences exceeding 200ms are detected using voice activity detection and removed. Each HE utterance is then globally time-stretched with PSOLA~\cite{charpentier1989pitch} to approximate the duration and speaking rate of its EL utterance. WEO features are extracted for both signals, and DTW is used to calculate the warping path between EL-HE feature sequences. The EL waveform is aligned to the HE waveform via frame-wise PSOLA along the warping path. Thus, potential PSOLA artifacts are limited to EL signals, while the HE signals remain mostly artifact-free as training targets. 

To maximize dataset variability, we did not restrict the alignment to one-to-one EL-HE pairs. Instead, each EL utterance is aligned with all available HE realizations of the same linguistic content across speakers. As a result, the number of aligned training pairs increased from 3,298 one-to-one to 19,592 EL-HE combinations.

\begin{table*}[h]
\centering
\small
\renewcommand{\arraystretch}{1.1}
\setlength{\tabcolsep}{8.2pt}
\setlength{\aboverulesep}{0pt}
\setlength{\belowrulesep}{0pt}
\caption{Objective evaluation (CER, ±95\% CI) on HE ground-truth (GT), EL, baselines (FreeVC, and XVC), and our model variants. Only significant improvements over \textit{w/o guided loss} are shown. Best results in \textbf{bold} (\textsuperscript{$\blacktriangle$} higher-is-better, \textsuperscript{$\blacktriangledown$} lower-is-better).}
\label{tab:objective}
\begin{tabular}{l c c ccc c c}
\toprule
Method & CER (\%)\textsuperscript{$\blacktriangledown$} (Whisper) & wvMOS \textsuperscript{$\blacktriangle$} &  SIG\textsuperscript{$\blacktriangle$}  & BAK\textsuperscript{$\blacktriangle$}  & OVRL\textsuperscript{$\blacktriangle$}  & SIM\textsuperscript{$\blacktriangle$} & Log-$F_{0}$ RMSE \textsuperscript{$\blacktriangledown$} \\
\midrule
GT & 2.9$\, \pm 1.1$ & 4.00 & 3.48 & 4.11 & 3.20 & 0.89 & - \\
EL & 88.2 $ \pm 51.2$ & -0.28 & 3.14 & 3.12 & 2.41 & 0.55 & 0.62 \\
\cmidrule(lr){1-8}
FreeVC  & 140.3 $ \pm 30.2$ & 3.52 & 3.27 & 3.99 & 2.91 & 0.71 & 0.40 \\
XVC     & 61.2 $ \pm 8.3$  & 3.59 & 3.32 & \textbf{4.02\textsuperscript{$\blacktriangle$}} & 3.00 & 0.63 & 0.37 \\
\cmidrule(lr){1-8}
\textit{Ours: w/o guided loss} & 53.7 $ \pm 6.0$ & 3.17 & 3.29 & 3.88 & 2.90 & 0.86 & 0.35 \\
\hspace{8mm}\textit{+WavLM}            & \textbf{40.9\textsuperscript{$\blacktriangledown$}}$ \pm 1.8$  & 3.26 & 3.32 & 3.93 & 2.94 & 0.84 & \textbf{0.34}\textsuperscript{$\blacktriangledown$} \\
\hspace{8mm}\textit{+WavLM+HF}        & \textbf{41.9\textsuperscript{$\blacktriangledown$}} $ \pm 1.8$  & 3.76 & \textbf{3.43\textsuperscript{$\blacktriangle$}} & \textbf{4.00\textsuperscript{$\blacktriangle$}} & \textbf{3.09\textsuperscript{$\blacktriangle$}} & 0.87 & \textbf{0.34}\textsuperscript{$\blacktriangledown$} \\
\hspace{8mm}\textit{+WavLM+HF+$F_{0}$}      & 46.7 $ \pm 3.9$  & 3.70 & 3.42 & 3.97  & 3.06 & 0.87 & 0.35 \\
\hspace{8mm}\textit{+WEO+HF}          & 44.9 $ \pm 1.7$  & 3.69 & 3.39 & \textbf{4.02\textsuperscript{$\blacktriangle$}}  & 3.05 & 0.86 & \textbf{0.34}\textsuperscript{$\blacktriangledown$} \\

\hspace{8mm}\textit{+WEO+WavLM+HF}     & 47.1 $ \pm 7.1$  & 3.70 & 3.38 & 3.98  & 3.02 & 0.87 & \textbf{0.34}\textsuperscript{$\blacktriangledown$} \\

\hspace{8mm}\textit{+BNF+HF}  & 55.41 $ \pm 1.8$  & \textbf{3.82}\textsuperscript{$\blacktriangle$} & 3.42 & 3.98  & 3.07 & \textbf{0.88}\textsuperscript{$\blacktriangle$} & 0.35 \\
\bottomrule
\end{tabular}
\vspace{-5mm}
\end{table*}

\subsection{Fine-tuning}

We do supervised fine-tuning using the parallel aligned ELHE database. EL speech serve as input and HE speech act as target.

Unlike the pretraining stage, the encoder is unfrozen during fine-tuning. While StreamVC reported that updating the encoder cause speaker leakage~\cite{yang2024streamvc}, this risk is mitigated by the acoustic differences between EL and HE speech.

In addition to the original loss functions used in StreamVC, we use several perceptually guided losses. The losses are: 
\textbf{(i) Perceptual loss (WavLM)} computed on WavLM~\cite{chen2022wavlm} features as introduced in FINALLY~\cite{babaev2024finally}; 
\textbf{(ii) Human feedback (HF) loss} inspired by FINALLY, defined as the negative mean UTMOS score~\cite{saeki2022utmos} (without PESQ); 
\textbf{(iii) $F_{0}$ contour loss} via log-$F_{0}$ MSE using pitch contours predicted with fast context-base pitch estimator\footnote{https://github.com/CNChTu/FCPE}(FCPE); 
\textbf{(iv) Intelligibility loss (BNF)} introduced in~\cite{zhou2023crosslingual} using MSE between pre-softmax Conformer-CTC~\cite{gulati2020conformer} bottleneck features (model fine-tuned on the HE portion of ELHE; features taken from the final layer before softmax); and \textbf{(v) Intelligibility loss (WEO)} using MSE between Whisper-small encoder hidden states (Whisper fine-tuned on the HE portion of the database).

To improve robustness, we use noise injection data augmentation for $30\%$ of the training samples. Random noise segments from a noise database~\cite{ko2017reverb} are added to the training samples with random signal-to-noise ratio (SNR) between 3 and 30~dB.  

Learning rates are set to $1.0e^{-4}$ (gen) and $1.25e^{-4}$ (disc) with $\beta_{1}{=}0.8$, $\beta_{2}{=}0.99$ and  cosine decay scheduling. The minimum rates are set to $2.5e^{-5}$ (gen) and $1.25e^{-5}$ (disc). Fine-tuning is performed for 150-250k steps with a batch size of 16. Guided losses were weighted to match the reconstruction loss, whereas the feature matching loss was up-weighted to highlight fine spectral details.
\vspace{-1mm}
\section{Evaluation}
\subsection{Objective Evaluation}
We first compare different model loss configurations objectively. The results are reported against the source EL and, where possible, to the GT. The evaluation is carried out on a set of 160 unmodified utterances (no PSOLA/DTW). In addition to our own variants, we use two offline state-of-the-art VC models as baselines: FreeVC~\cite{FreeVC2023}, and XVC~\cite{consistencyVC2023}. Since no standardized open-source EL-HE VC baselines exist, these models serve as reference points for indirect comparison and are fine-tuned on the HE portion of the ELHE corpus before one-shot evaluated on EL inputs. We report results across the objective metrics: \textbf{(i) Character error rate (CER)}: Transcriptions are done using zero-shot Whisper-small. CER reflects intelligibility. \textbf{(ii) DNSMOS}~\cite{reddy2021dnsmos}: Objective quality predictor model for speech quality \textbf{(SIG)}, background noise \textbf{(BAK)}, and overall perceptual quality \textbf{(OVRL)}. \textbf{(iii) wvMOS}~\cite{andreev2023hifipp}: MOS score predicted by wav2vec2.0. \textbf{(iv) Speaker similarity (SIM)}: Speaker similarity calculated via Resemblyzer~\cite{wan2018gee} between converted and target HE speech. \textbf{(vi) Log-$\bm{F_{0}}$ RMSE}: Difference of pitch contour between conversions and target HE speech. Additionally, we assess robustness under noisy conditions. Random quasi-stationary noise (wind, fan, street traffic, train, waterfalls) and non-stationary (background conversations, sirens, smartphone ringing, in-car sirens) are added to the EL input at varying SNR (0-25dB) levels. CER is computed on the VC speech.

\begin{figure}[t]
\begin{minipage}[b]{.5\linewidth}
  \centering
  \centerline{\includegraphics[width=1.024\linewidth]{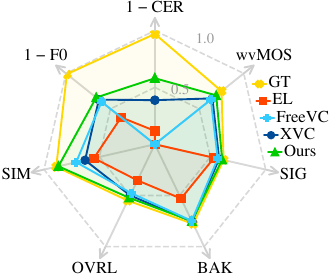}}
  \vspace{1.75mm}
  \centerline{(a) Objective results.}\medskip
\end{minipage}
\hspace{-0.024\linewidth}
\begin{minipage}[b]{0.5\linewidth}
  \centering
  \centerline{\includegraphics[width=0.97\linewidth]{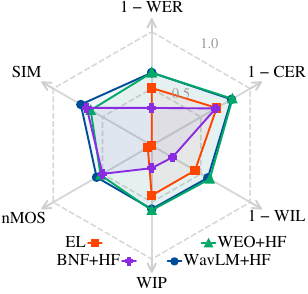}}
 
  \centerline{(b) Subjective results. }\medskip
\end{minipage}
\vspace{-9mm}
\caption{Normalized (0-1) evaluation metrics (higher is better). (a) Objective metrics comparing GT, EL, baselines (FreeVC \& XVC), and our best method (\textit{+WavLM+HF}). (b) Subjective metrics comparing EL vs. our method with three loss configurations (\textit{+WavLM+HF}, \textit{+WEO+HF}, and \textit{+BNF+HF}).}
\label{fig:res}
\vspace{-5mm}
\end{figure}

\begin{table}[t]
\centering
\small
\setlength{\tabcolsep}{1pt}
\caption{Subjective evaluation. EL vs. our method with three loss configurations (\textit{+WavLM+HF}, \textit{+WEO+HF}, and \textit{+BNF+HF}), evaluated on WER, CER, nMOS, and SIM.}
\label{tab:listening_overall}
\renewcommand{\arraystretch}{0.95}
\begin{tabular}{@{}lcccc@{}}
\toprule
Method & WER (\%)\textsuperscript{$\blacktriangledown$} & CER (\%)\textsuperscript{$\blacktriangledown$} & nMOS (1--5)\textsuperscript{$\blacktriangle$} & SIM (0--1)\textsuperscript{$\blacktriangle$} \\
\midrule
EL    & $49.4 \pm 19$ & $33.6 \pm 16$ & $1.1 \pm 0.2$ & — \\
\textit{+WavLM+HF} & \textbf{36.4}\textsuperscript{$\blacktriangledown$} $\pm$ 17 & $19.9 \pm 10$ & \textbf{3.3}\textsuperscript{$\blacktriangle$}  $\pm$ 0.4 & \textbf{0.77}\textsuperscript{$\blacktriangle$} $\pm$ 0.09 \\
\textit{+WEO+HF}  & \textbf{36.4}\textsuperscript{$\blacktriangledown$} $\pm$ 16 & \textbf{18.4}\textsuperscript{$\blacktriangledown$} $\pm$ 9 & $3.1 \pm 0.3$ & $0.68 \pm 0.09$ \\
\textit{+BNF+HF}  & $68.4 \pm 14$ & $35.2 \pm 11$ & $3.0 \pm 0.5$ & $0.71 \pm 0.07$ \\
\bottomrule
\end{tabular}
\vspace{-5mm}
\end{table}

\subsection{Subjective Evaluation}
 A listening survey was created using Go Listen~\cite{zhang2021golisten}. Three model configurations (\textit{+WavLM+HF}, \textit{+WEO+HF}, and \textit{+BNF+HF}) were selected for subjective evaluation and compared against the EL input. 22~participants took part in the survey. Each participant completed three tasks: \textbf{(1) orthographic transcription}, for which word error rate \textbf{(WER)}, \textbf{CER}, word information preserved \textbf{(WIP)}, and word information lost \textbf{(WIL)}~\cite{morris04_interspeech} were calculated; \textbf{(2) naturalness MOS (nMOS) rating}, using a 6-point scale (1 = not natural/very synthetic, 6 = completely natural/authentic), normalized to 1–5; and \textbf{(3) speaker similarity rating (SIM)}, using a 6-point scale (1 = definitely not the same speaker, 6 = definitely the same speaker), normalized to a 0–1 range. Four test sets (A–D) were created, each containing four EL utterances for transcription, five for naturalness, and three for similarity, yielding 12 unique EL signals with corresponding conversions into 2 female and 2 male voices. Audio samples were randomly drawn from the test set, with each method receiving an equal number of ratings across conditions.

\begin{figure}[t]
\vspace{-0.5mm}
  \centering
    \includegraphics[width=0.95\linewidth]{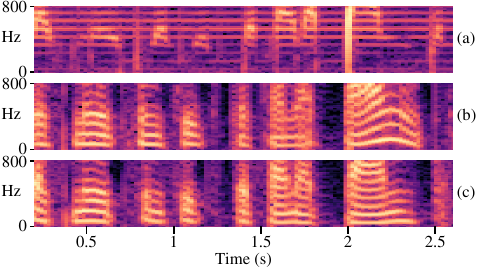}
    \vspace{-3mm}
  \caption{Mel-spectogram comparison: (a) EL input, (b) GT speech and (c) VC for model configuration \textit{+WavLM+HF}.}
  \label{fig:MELcomp}
  \vspace{-5mm}
\end{figure}

\vspace{-1mm}
\section{Results \& Discussion}

\textbf{Objective results (Table~\ref{tab:objective}, Fig.~\ref{fig:res}(a)).} The overall best objective performance is achieved with \textit{+WavLM+HF}, with \textit{+WEO+HF} performing similarly well, while \textit{+BNF+HF} improves wvMOS and SIM but clearly degrades CER. A general trend is that adding more than 2 auxiliary losses to the GAN setup does not help. The best results come from combining a single perceptual loss (WavLM or WEO) with one speech quality loss (HF). Additional terms (e.g., $F_0$ or further intelligibility losses) tend to reduce both intelligibility and quality, likely due to competing gradients preventing convergence. The different losses emphasize distinct properties. WavLM and WEO losses encourage noise-robust phonetic representations that suppress EL-specific buzzing artifacts, improving intelligibility (CER) at the cost of slightly lower perceptual quality scores (DNSMOS, wvMOS). In contrast, the BNF loss, is highly content-constrained and sensitive to EL–HE misalignment, which can over-penalize local timing errors and lead to unstable or “mumbled” realizations, explaining the reduced intelligibility despite gains in naturalness and speaker similarity.

In Fig.~\ref{fig:res}(a) and (b), all metrics are normalized (0-1), with error-type measures (e.g., WER, CER, WIL, and $F_{0}$ RMSE) inverted as $1{-}$metric so that larger values indicate better performance (i.e., larger areas in spider plots map to better results). In Fig.~\ref{fig:res}(a), our \textit{+WavLM+HF} variant is superior to EL and the models FreeVC and XVC in all metrics. Performance even approaches the GT in nearly all metrics except for $F_{0}$ RMSE and CER, highlighting prosody and intelligibility as the main directions for future research.

\textbf{Subjective results (Table~\ref{tab:listening_overall}, Fig.~\ref{fig:res}(b)).}
The listening tests align with the objective trends. \textit{+WavLM+HF} and \textit{+WEO+HF} both reduce WER and CER and improve nMOS versus EL. \textit{+WavLM+HF} attains the best nMOS and highest SIM among the two. \textit{+BNF+HF} improves nMOS and SIM but reduces WER and CER, mirroring Table~\ref{tab:objective}. Fig.~\ref{fig:res}(b) shows balanced improvements from \textit{+WavLM+HF} and \textit{+WEO+HF} across all metrics, with particularly strong gains in nMOS and smaller but consistent improvements in all four intelligibility measures (WER, CER, WIP, WIL). By contrast, \textit{+BNF+HF} shifts towards high nMOS and SIM, but achieves low intelligibility.

\textbf{Time–frequency analysis (Fig.~\ref{fig:MELcomp}).}
The plot shows (a) EL source, (b) HE ground-truth, and (c) conversion into the same HE target. The conversion closely matches the ground-truth. The constant $F_{0}$ of the EL signal and the mechanical buzzer noise are absent, replaced by a more natural harmonic structure with clear formants. A low noise floor remains visible, but this is also present in the HE spectra. The $F_{0}$ contour shows natural variability with aligned harmonics. The conversion already achieves a close approximation of HE speech. Only the strong pitch excursion at the sentence ending appears flattened. This suggest that prosody modeling still leaves room for improvement. 

\begin{figure}[t]
\vspace{-0.5mm}
  \centering
    \includegraphics[width=0.95\linewidth]{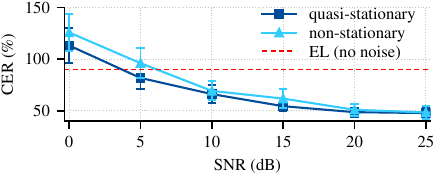}
    \vspace{-3mm}
  \caption{CER as a function of SNR (0-25dB) for conversions under quasi-stationary and non-stationary noise conditions. The red line shows the CER obtained from unconverted, noise-free EL speech.}
  \label{fig:CERvsSNR}
  \vspace{-5mm}
\end{figure}

\textbf{Noise robustness (Fig.~\ref{fig:CERvsSNR}).}
Under noisy conditions, CER degrades as SNR decreases for both quasi-stationary and non-stationary conditions. Non-stationary noise is more harmful due to rapid spectral changes that may confuse the content encoder. At moderate to high SNRs, clear gains in terms of intelligibility over EL can be observed. However, as SNR decreases, the performance gap narrows, and below 5dB SNR the converted speech shows lower intelligibility than the unconverted EL input.

\vspace{-1mm}
\section{Conclusion \& Future Work}

We demonstrated how the lightweight, fully convolutional StreamVC model can be adapted to EL speech by integrating Whisper-based time alignment in preprocessing and perceptual guided losses in training, leading to improved harmonic restoration, intelligibility and naturalness despite its simple architecture. Objective and subjective results show that the best loss configuration (\textit{+WavLM+HF}) consistently outperforms EL speech, baseline VC models, other loss configurations, and narrows the gap to HE ground-truth performance even under noisy conditions. Only prosody and intelligibility remain limited and are key future perspectives for EL speech rehabilitation.

To address these bottlenecks, we plan two future research directions. First, methodologies such as context-aware Transformer decoders or expressive synthesis could enhance prosody and intelligibility. Second, we plan to realize a complete streaming pipeline and evaluate its latency on CPU-based embedded devices, targeting real-time EL–to-HE VC on constrained hardware.

\vspace{-1mm}
\section{ACKNOWLEDGMENTS}
\vspace{-1mm}
We thank all participants of the listening survey for their contributions. This research was funded in part by the Austrian Science Fund (FWF) [10.55776/PAT5948223] and utilized the Austrian Scientific Computing (ASC) infrastructure. ChatGPT (OpenAI) was used for research, visualization, and proofreading.


\bibliographystyle{IEEEbib}
\bibliography{strings,refs}

\end{document}